\documentclass[twocolumn]{aastex631}

\received{December 14, 2021}
\revised{Jan 19, 2022}
\accepted{Jan 23, 2022}

\shorttitle{Hard-state optical wind in MAXI~J1803$-$298}
\shortauthors{Mata S\'anchez et al.}
\graphicspath{{./}{figures/}}

\begin{document}

\title{Hard-state optical wind during the discovery outburst of the black-hole X-ray dipper MAXI~J1803$-$298}

\correspondingauthor{D. Mata S\'anchez }
\email{matasanchez.astronomy@gmail.com, dmata@iac.es}

\author[0000-0003-0245-9424]{D. Mata S\'anchez}
\affiliation{Instituto de Astrof\'isica de Canarias, E-38205 La Laguna, Tenerife, Spain}
\affiliation{Departamento de Astrof\'isica, Univ. de La Laguna, E-38206 La Laguna, Tenerife, Spain}

\author[0000-0002-3348-4035]{T. Mu\~noz-Darias}
\affiliation{Instituto de Astrof\'isica de Canarias, E-38205 La Laguna, Tenerife, Spain}
\affiliation{Departamento de Astrof\'isica, Univ. de La Laguna, E-38206 La Laguna, Tenerife, Spain}

\author[0000-0002-1813-9137]{V. A. C\'uneo}
\affiliation{Instituto de Astrof\'isica de Canarias, E-38205 La Laguna, Tenerife, Spain}
\affiliation{Departamento de Astrof\'isica, Univ. de La Laguna, E-38206 La Laguna, Tenerife, Spain}

\author[0000-0002-4344-7334]{M. Armas Padilla}
\affiliation{Instituto de Astrof\'isica de Canarias, E-38205 La Laguna, Tenerife, Spain}
\affiliation{Departamento de Astrof\'isica, Univ. de La Laguna, E-38206 La Laguna, Tenerife, Spain}

\author[0000-0003-2276-4231]{J. S\'anchez-Sierras}
\affiliation{Instituto de Astrof\'isica de Canarias, E-38205 La Laguna, Tenerife, Spain}
\affiliation{Departamento de Astrof\'isica, Univ. de La Laguna, E-38206 La Laguna, Tenerife, Spain}

\author[0000-0003-1513-1460]{G. Panizo-Espinar}
\affiliation{Instituto de Astrof\'isica de Canarias, E-38205 La Laguna, Tenerife, Spain}
\affiliation{Departamento de Astrof\'isica, Univ. de La Laguna, E-38206 La Laguna, Tenerife, Spain}

\author[0000-0001-5031-0128]{J. Casares}
\affiliation{Instituto de Astrof\'isica de Canarias, E-38205 La Laguna, Tenerife, Spain}
\affiliation{Departamento de Astrof\'isica, Univ. de La Laguna, E-38206 La Laguna, Tenerife, Spain}

\author[0000-0003-1038-9104]{J. M. Corral-Santana}
\affiliation{European Southern Observatory, Alonso de C\'ordova 3107, Vitacura, Casilla 19001, Santiago de Chile, Chile}

\author[0000-0002-5297-2683]{M. A. P. Torres}
\affiliation{Instituto de Astrof\'isica de Canarias, E-38205 La Laguna, Tenerife, Spain}
\affiliation{Departamento de Astrof\'isica, Univ. de La Laguna, E-38206 La Laguna, Tenerife, Spain}



\begin{abstract}
We present twelve epochs of optical spectroscopy taken across the discovery outburst of the black hole candidate MAXI~J1803$-$298 with the GTC and VLT telescopes. The source followed a standard outburst evolution with hard and soft states. The system displays a triangular shape in the hardness intensity diagram, consistent with that seen in high inclination black hole transients and the previously reported detection of X-ray dips. The two epochs observed during the initial hard state exhibited asymmetric emission line profiles, including a P-Cygni profile simultaneously detected in H$\rm \alpha$ and He \textsc{i} 6678, which indicates the presence of an optical wind in the system. The remaining spectra, obtained during the transition to the soft state and the subsequent decay, are instead characterized by narrower, double peaked emission lines embedded into broad absorption components. One epoch (intermediate state) also includes near-infrared coverage, revealing complex line profiles in the Paschen and Bracket series, which suggests that the outflow is still present during the outburst decay through the soft state. The growing list of low-mass X-ray binaries with optical and near-infrared outflow signatures indicates that these are common features. Furthermore, the lowest luminosity spectrum exhibits an H$\rm \alpha$ full-width-at-half-maximum of $1570\pm 100 \, {\rm km\, s^{-1}}$. This, together with previous constraints on the binary parameters, allows us to favor a compact object mass of $\sim 3-10\, M_{\rm \odot}$, further supporting its black hole nature.

\end{abstract}



\section{Introduction} \label{sec:intro}

Low-mass X-ray binaries (LMXBs) consist of either a neutron star or a black hole (BH) accreting matter from a Roche-Lobe filling stellar companion ($\leq 1\,M_{\rm \odot}$) via an accretion disk. The majority of LMXBs are transient systems that spend most of their lives in a faint, quiescent state ($L_{\rm X}\sim 10^{31-33} \, {\rm erg\, s^{-1}}$), but exhibit short-lived outbursts with heavily enhanced emission ($L_{\rm X}\sim 10^{36-38} \, {\rm erg\, s^{-1}}$). More than $\sim$ 60 LMXBs are thought to harbor a BH, with only 21 of them confirmed by dynamical studies (for a review see \citealt{Casares2014,Corral-Santana2016}; also \citealt{MataSanchez2015b,Torres2021}). 

During outburst, BH transients show two characteristic X-ray states, hard and soft, as well as outflows in the form of jets and winds (e.g. \citealt{Fender2004, Fender2016}). In particular, accretion disk winds were initially discovered in X-rays as blue-shifted absorptions in transitions of highly ionized species (e.g. \citealt{Ueda1998, Miller2006}). These typically appear during the soft state, and are solely detected in high inclination systems \citep{Neilsen2009, Ponti2012, DiazTrigo2016}. In the optical range, lower ionization winds have been found in a number of BH transients over the last few years. These are detected as P-Cygni profiles, absorption troughs and broad emission line wings in reference lines such as H$\rm \alpha$ and He~\textsc{i} $5876~\rm{\AA}$ (e.g. \citealt{Munoz-Darias2016, Munoz-Darias2019}). 

MAXI~J1803$-$298 (hereafter J1803) is a BH candidate discovered on May 1st 2021 \citep{Serino2021}. During a $\sim$ 7-month long outburst, it followed the classic X-ray evolution, with hard and soft states. The detection of periodic dips (lasting $\sim 5000\, {\rm s}$) during the initial stages of the outburst indicates a high orbital inclination and likely orbital period of $P_{\rm orb}\sim 7\, {\rm h}$ \citep{Gendreau2012,Xu2021}. Interestingly, preliminary reports suggested the presence of wind-related features in both the X-ray and optical spectra of the source \citep{Miller2021, Buckley2021}.  

In this work, we present multi-epoch optical and near-infrared spectroscopic observations of J1803, together with an X-ray monitoring of the outburst.

\begin{deluxetable*}{cccccccccccc}
\tablenum{1}
\tablecaption{Journal of observations.\label{tab:obslog}}
\tablewidth{0pt}
\tablehead{
\colhead{Epoch} & \colhead{Date} & \colhead{TST$^{\rm a}$}  & \colhead{Grism/slit} &\colhead{R} & \colhead{\#} & \colhead{$\rm {T_{exp}}^{\rm b}$} & \colhead{Seeing$^{\rm c}$}  & \colhead{g-band} &  \colhead{r-band} &  \colhead{X-ray state$^{\rm d}$} &  \colhead{$\rm H\alpha$ FWHM$^{\rm e}$}\\
\nocolhead{} &\colhead{(dd/mm)} & \colhead{(d)}& \colhead{(/\arcsec)} &\nocolhead{} & \nocolhead{} & \colhead{(s)} & \colhead{(\arcsec)}  & \colhead{(mag)}  & \colhead{(mag)} & &\colhead{($\rm km\, s^{-1}$)} 
}
\startdata
GTC-1& 03/05 & 2.179 & R1000B/1.0 & 833 & 8 & 300 & 1.7& $16.06\pm 0.01$ & & Hard & $1360\pm 210$\\
GTC-2& 05/05 & 4.189  & R2500R/1.0 & 2143  & 6 & 450 & 1.6& $16.06\pm 0.01$& $16.23\pm 0.01$ & Hard& $1280\pm 70$\\
GTC-3& 07/05 & 6.194 & R2500R/1.0 & 2143 & 4 & 600 & 1.1& $16.08\pm 0.01$  &  & Hard/HIMS& $1240\pm 290$\\
GTC-4& 09/05 & 8.182 & R2500R/1.0  & 2143 & 4 & 600& 1.1& $16.16\pm 0.01$ &  & Hard/HIMS& $1270\pm 270$\\
GTC-5& 15/05 & 14.137  & R2500R/1.0 & 2143 & 6 & 450 & 3.4& $16.22\pm 0.03$&  & HIMS& $1070\pm 230$\\
VLT-1& 15/05 & 14.264   & UVB/1.0 & 5400 & 1 & 581 & 0.82 & &  & HIMS\\
 &   &    & VIS/0.9 &  8900 & 1 & 554 & & &  &  &$1020\pm 70$\\
 &   &    & NIR/0.9 &  5600 & 4 & $ 47$ & & &  & \\
GTC-6& 16/05 & 15.119  & R2500R/1.0 & 2143  & 6 & 450 & 2.0& $16.36\pm 0.01$ &  & HIMS& $890\pm 80$\\
GTC-7& 22/05 & 21.178  & R2500R/1.0 & 2143 & 4 & 600 & 1.9& $16.43\pm 0.01$ &  & SIMS & $1300\pm 230$\\
GTC-8& 02/06 & 32.153  & R2500R/1.0 & 2143 & 4 & 600 & 1.7& $16.82\pm 0.02$ &  & Soft & $1240\pm 300$\\
GTC-9& 16/06 & 46.053  & R2500R/1.0 & 2143 & 4 & 600 & 1.2&  & $17.02\pm 0.03$   & Soft& $1470\pm 670$\\
GTC-10& 02/07 & 62.050 & R2500R/1.0  & 2143& 4 & 600 & 1.6& $17.40\pm 0.03$ &  & Soft& $1180\pm 280$\\
GTC-11& 02/08 & 93.943 & R2500R/1.0 & 2143  & 9 & 600 & 1.4& $17.74\pm 0.03$ & & Soft & $1570\pm 100$\\
\enddata
\tablecomments{a) TST stands for ``time since trigger'', defined as the mid-exposure \rm{MJD} from the average spectrum of each epoch referred to the outburst detection trigger ($\rm{MJD}\, 59336.0$). b) The exposure time corresponds to the individual exposures within each epoch. c) The seeing value was measured on the g-band acquisition images from each GTC epoch, except for GTC-9, where only r-band acquisition images were available. The seeing value for VLT-1 observations corresponds to the measured image quality. d) The X-ray state associated with each observation was classified attending to the hardness intensity diagram (Fig. \ref{fig:HID}) and the QPOs detection. e) Full-width-at-half-maximum of a Gaussian fit to the $\rm H\alpha$ line, as described in Sec.\ref{sec:analysis}}
\end{deluxetable*}

\section{Observations} \label{sec:observations}

The spectroscopic database covers twelve different epochs across the outburst. A standard reduction of the available acquisition images allowed us to derive contemporaneous photometry, which we report as part of the observing log (Table \ref{tab:obslog}).

We obtained 11 epochs (59 spectra) with the \mbox{10.4-m} Gran Telescopio Canarias (GTC) at the Roque de los Muchachos Observatory (La Palma, Spain), equipped with the Optical System for Imaging and low-Intermediate-Resolution Integrated Spectroscopy (OSIRIS, \citealt{Cepa2000}). All observations used a $1\arcsec $ slit width. The R1000B grism ($R=833$ and $3600 - 7800\, {\rm \AA}$ coverage) was used during the first epoch, and the R2500R grism ($R=2143$ and $5600 - 7700\, {\rm \AA}$ coverage) for the remaining observations. We reduced the spectra following standard procedures making use of semi-automatic routines based on \textsc{iraf}\footnote{IRAF is distributed by the National Optical Astronomy Observatories, operated by the Association of Universities for Research in Astronomy, Inc., under contract with the National Science Foundation.} (via \textsc{pyraf}, the python implementation maintained by the community) and \textsc{molly}\footnote{MOLLY software developed by T. R. Marsh.} tasks.

An additional epoch was observed with the Very Large Telescope (VLT; Paranal Observatory, Chile) equipped with the spectrograph X-Shooter \citep{Vernet2011}. We employed slit widths of $1\arcsec $, $0.9\arcsec $ and $0.9\arcsec $ for the UVB, VIS and NIR arms, which produced spectral resolutions of $R=5400$, $R=8900$ and $R=5600$, respectively. The reduction was performed with the latest available version of the official pipeline, and allowed us to obtain a single spectrum covering the wavelength range of $3000 - 24800\, {\rm \AA}$.

\section{Analysis and Results} \label{sec:analysis}

\subsection{Hardness Intensity Diagram} \label{sec:hid}

\begin{figure}[ht!]
\includegraphics[keepaspectratio, trim=0cm 0cm -0.5cm 0cm, clip=true, width=0.5\textwidth]{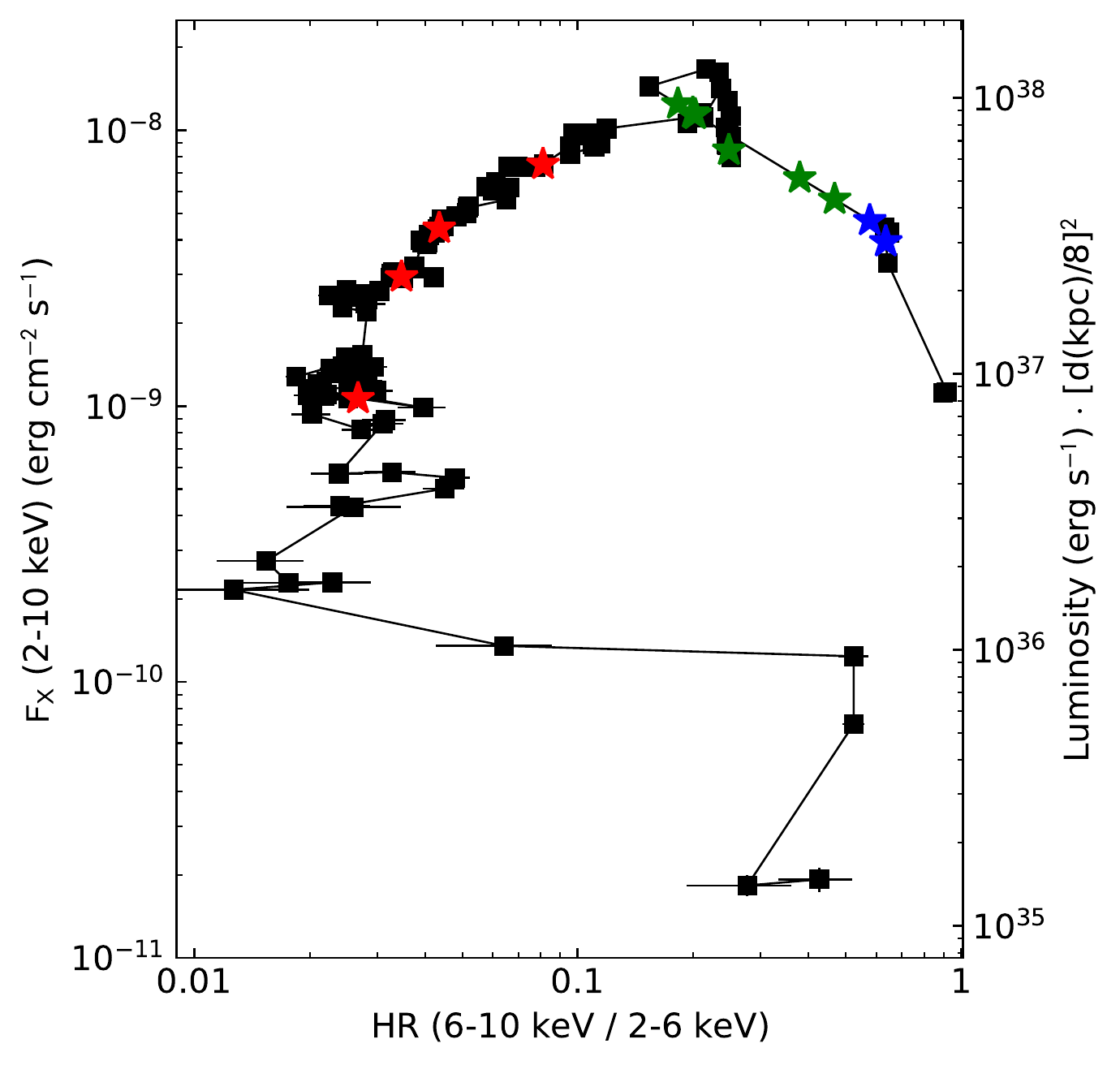}
\caption{HID for the J1803 outburst as seen by \textit{NICER} (black squares). Note the triangular shape and the ``plume'' at $F_{\rm{X}} (2-10\, \rm{keV})\sim 10^{-8}\,{\rm erg\, cm^{-2}\,s^{-1}}$ and $\rm{HR}\sim 0.2$, characteristic of high-inclination BH transients. We use stars to mark the spectroscopic epochs, with blue color for the hard state, green for the intermediate states, and red for the soft state. \label{fig:HID}}
\end{figure}

In order to determine the X-ray state of J1803 during the optical observations, we analyzed the publicly available data from the \textit{NICER} instrument (Neutron star Interior Composition Explorer; \citealt{Gendreau2012}), following a standard prescription for the data reduction\footnote{\url{https://heasarc.gsfc.nasa.gov/docs/nicer/analysis_threads/}} (see e.g., \citealt{Cuneo2020a}). We carried out a phenomenological fit of the observations between days MJD~59336 and MJD~59511 in XSPEC (v.12.12.0, \citealt{Arnaud1996}) using a multicolor disk (DISKBB; \citealt{Mitsuda1984}) plus a Comptonization component (NTHCOMP; \citealt{Zdziarski1996}, \citealt{Zycki1999}). We used the Tuebingen–Boulder interstellar medium absorption model (TBABS in XSPEC) with cross-sections of \citet{Verner1996} and abundances of \citet{Wilms2000} to account for interstellar absorption. We assumed an equivalent hydrogen column of $N_{\rm H}=3.2\times 10^{21}\, \rm{cm}^{-2}$ \citep{Bult2021}. From the fits, we derived fluxes in the $2-6$ keV and $6-10$ keV energy bands to construct the hardness intensity diagram (HID; \citealt{Homan2001}). This allows for a direct comparison with the BH sample presented in \citet{Munoz-Darias2013}. Since the distance is unknown, we assumed a standard value of 8 kpc to compute the X-ray luminosity in Fig. \ref{fig:HID}. We note that a detailed X-ray study of the \textsc{NICER} data is beyond the scope of this Letter. 

The resulting HID (Fig. \ref{fig:HID}) shows a q-shaped hysteresis pattern, a behavior typically seen in BH (e.g. \citealt{Fender2004}) and neutron star \citep{Munoz-Darias2014} LMXBs. The HID has triangular shape as well as a spike during the hard-to-soft transition (a.k.a. plume), both distinctive features of high-inclination BH transients (\citealt{Munoz-Darias2013}). This is in agreement with the detection of strong X-ray dips, which imply $\rm{i}>60\deg$ \citep{Frank1987}. We note that the hardness value reached during the soft state is similar to that shown by some mid-to-low inclination BHs in \citet{Munoz-Darias2013}. However, the latter study uses RXTE data, whose high-energy coverage ($>$10 keV) and sensitivity enables an accurate modelling of the Comptonization component. This is not the case for the soft state observations used here, whose hard X-ray band fluxes are most likely underestimated due to \textit{NICER} lower sensitivity at high energies. Thus, the actual value of the X-ray color during the soft state has to be taken with caution.

Fig. \ref{fig:HID} shows the location of the spectroscopic epochs in the HID. GTC-1 and GTC-2 (see Table \ref{tab:obslog}) were obtained during the raise of the outburst in the hard state. Epochs GTC-3 to GTC-6 (including VLT-1) were obtained during either the bright hard state or the hard-intermediate state (HIMS), as shown by the detection of type-C quasi-periodic oscillations (QPOs; \citealt{Xu2021,Chand2021}). The discovery of type-B QPOs coincident with epoch GTC-7 \citep{Ubach2021} is a defining trait of a soft-intermediate state (SIMS, e.g. \citealt{Belloni2011}). The remaining spectra correspond to the soft state.

\subsection{Evolution of the optical spectra}

\begin{figure*}[ht!]
\includegraphics[keepaspectratio, trim=0cm 1cm 0cm 0cm, clip=true, width=0.85\textwidth]{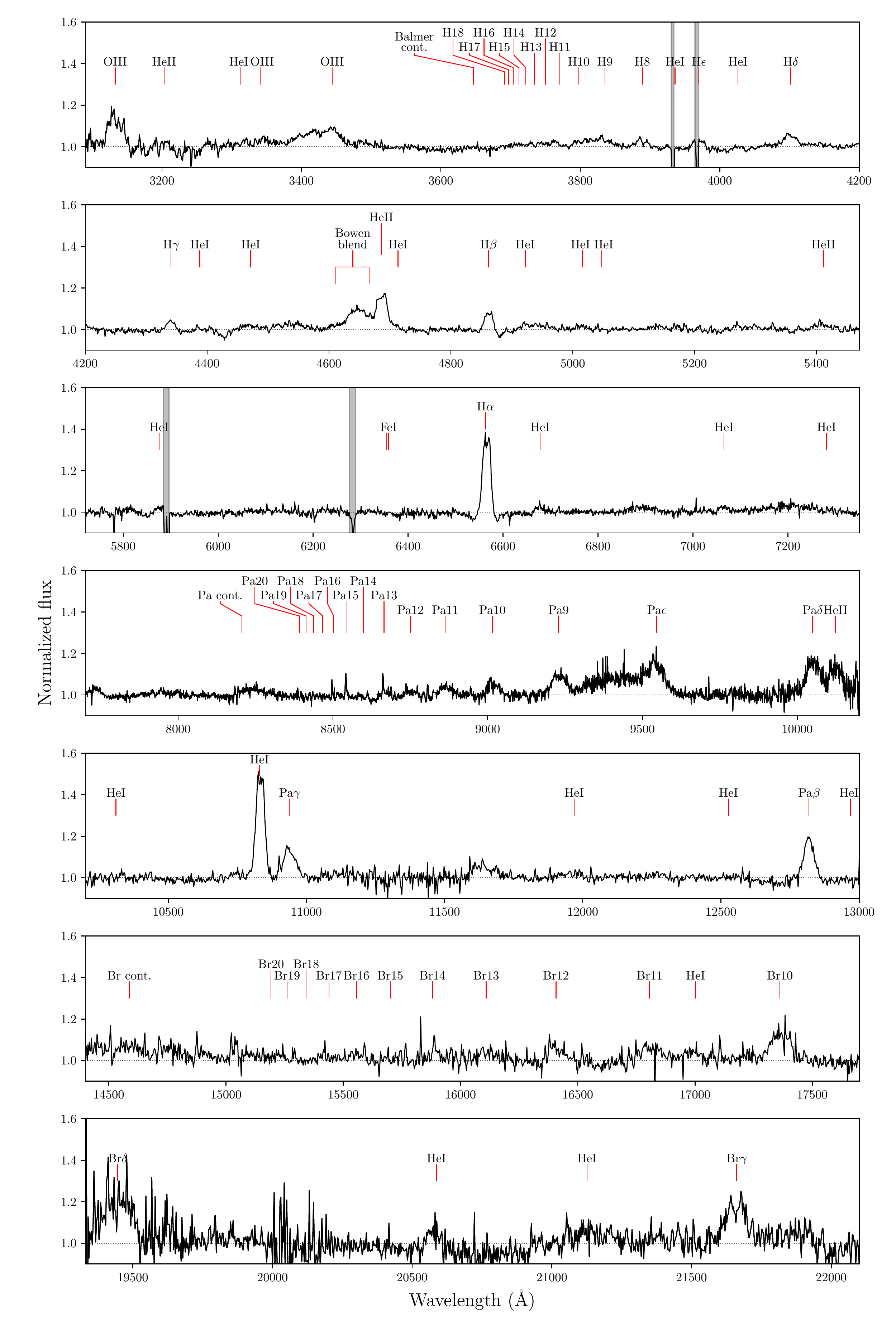}
\caption{Normalized spectrum of the VLT-1 epoch. Transitions expected in LMXBs are marked thorough the spectrum, including the hydrogen Balmer, Paschen and Bracket series; He \textsc{i} and \textsc{ii}; and fluorescence lines (e.g. Bowen blend). The most prominent DIBs and subtraction residuals are marked as shaded regions.} \label{fig:xshooter}
\end{figure*}

\begin{figure*}[ht!]
\includegraphics[keepaspectratio, trim=0cm 0cm 0cm 0cm, clip=true, width=\textwidth]{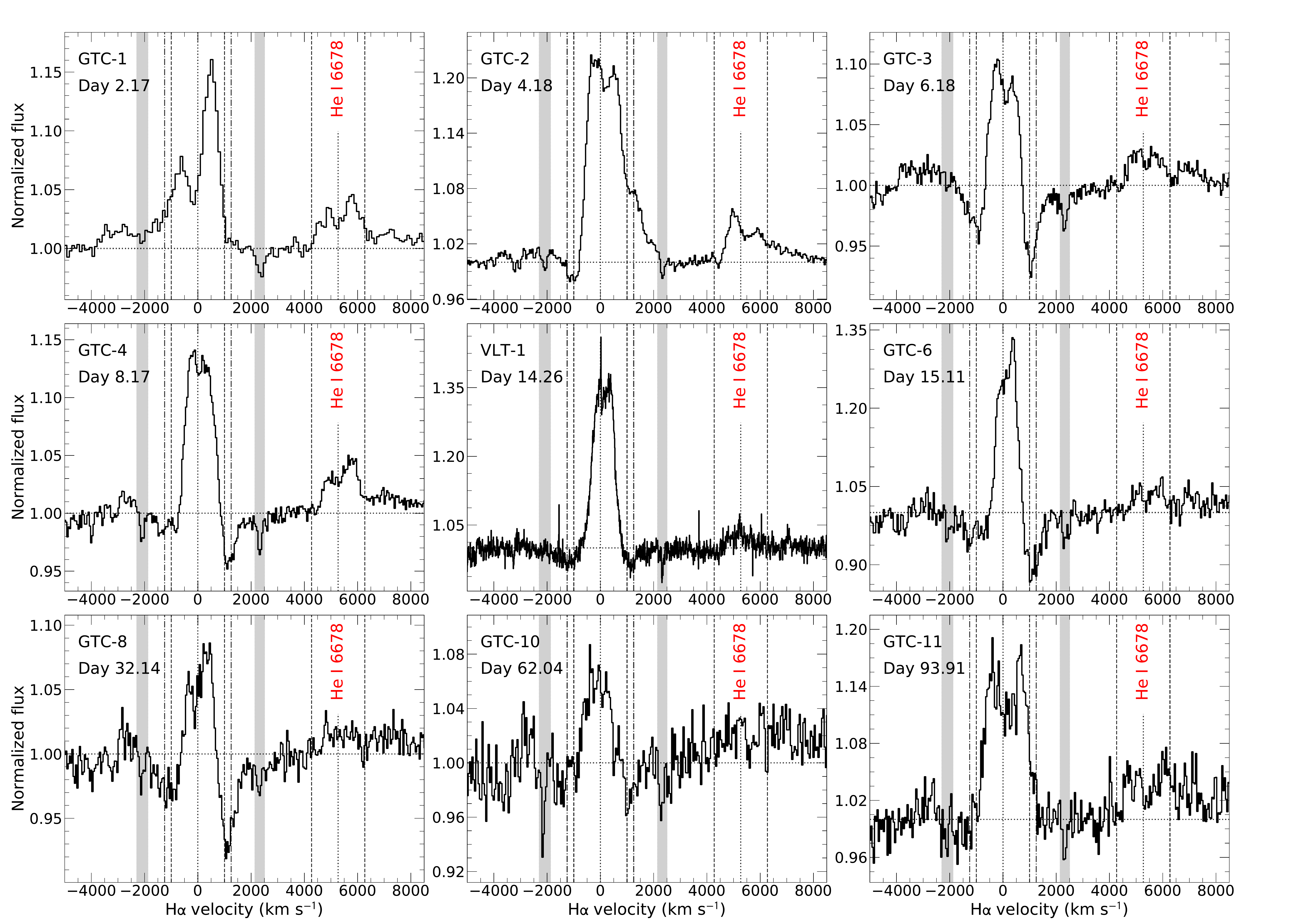} 
\caption{Normalized spectra from 9 of the observed epochs centered around the H$\alpha$ line, where the x-axis shows the velocity offset with respect to the wavelength of rest (vertical, dotted line). Vertical, dashed lines mark the velocity at values $\pm 1000\, \rm{km\, s^{-1}}$ while dash-dotted lines do at $\pm 1250\, \rm{km\, s^{-1}}$, which correspond to remarkable features associated with outflows in the different epochs. The nearby line He \textsc{i} 6678 is also shown, with vertical, dashed lines marking the $\pm 1000\, \rm{km\, s^{-1}}$ velocity relative to this line rest wavelength (vertical, dotted line). Telluric bands and DIBs are depicted as shaded regions. \label{fig:inset}}
\end{figure*}

\begin{figure*}[ht!]
\includegraphics[keepaspectratio, trim=0cm 1cm 0cm 0cm, clip=true, width=\textwidth]{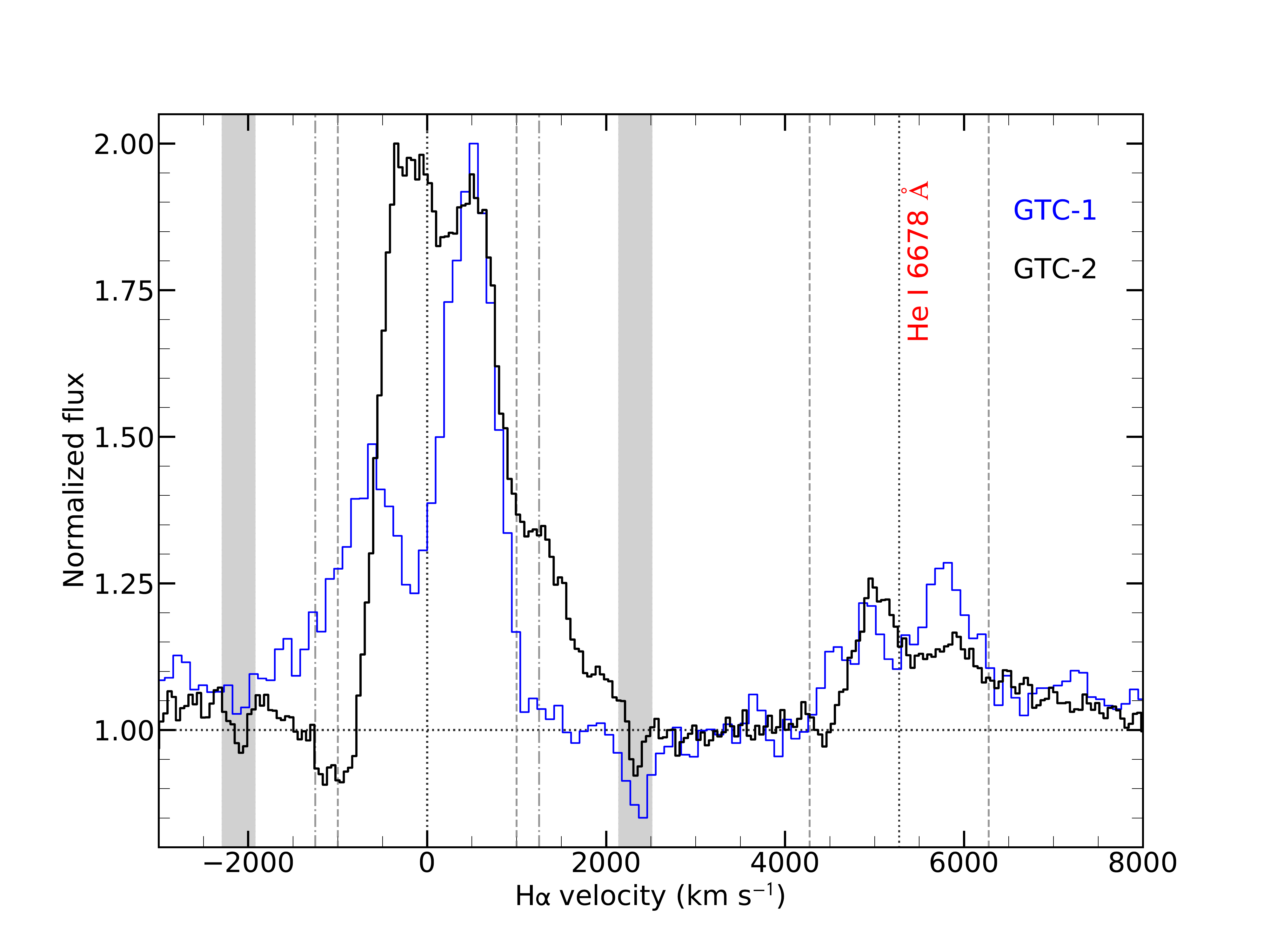}
\caption{Normalized spectrum of the GTC-1 epoch (blue) and GTC-2 epoch (black), both centered at H$\rm \alpha$. Telluric bands, DIBs and reference velocities have been marked as in Fig. \ref{fig:inset}. \label{fig:overplot}}
\end{figure*}

We normalized the spectra to their continuum level. We used different masks (in order to avoid emission and telluric lines), explored a range of polynomial orders, and normalized both the full spectra and the particular regions of interest separately. While these resulted in a reliable normalization (see e.g. Fig. \ref{fig:xshooter}, \ref{fig:inset}), we note that the continuum shape was not smooth. The presence of relatively deep, known diffuse interstellar bands (DIBs, such as $\sim 6613\rm{\AA}$) and other unidentified features hampered a better characterization of the spectral continuum, although we stress that this has no impact on the results on this paper.

The VLT-1 spectrum covers the widest wavelength range, from near-ultraviolet to the near-infrared (Fig. \ref{fig:xshooter}). We identify hydrogen emission lines corresponding to the Balmer, Paschen and Bracket series. He \textsc{i} and \textsc{ii} transitions are also detected in emission, with the strongest examples at $10830.17\rm{\AA}$ and $4685.75\rm{\AA}$, respectively. Additionally, high excitation lines produced by fluorescence, such as the Bowen blend (composed of a mixture of C, O and N lines), are also identified. 

The GTC spectra, while covering a more limited wavelength range, allow us instead to analyze the evolution of line profiles during the outburst. In particular, we focus our analysis on H$\rm \alpha$, a reference line present during the whole event, and also discuss the profile of the adjacent He \textsc{i} $6678.15\rm{\AA}$ line when detected. In order to provide quantitative results on the profile variability, we perform a multi-Gaussian fit to H$\rm \alpha$. We fit to each spectrum two emission Gaussians of the same width ($W$), with variable heights ($H_{\rm blue}$ and $H_{\rm red}$) and centroid positions, which are conveniently defined in terms of the peak-to-peak separation (DP) and the centroid of the double-peaked structure ($\gamma$). Due to the appearance of broad, absorption components during certain epochs, we also included when necessary a third Gaussian in absorption. All reported uncertainties correspond to a statistical $1\sigma$ value.

We find dramatic changes in the H$\rm \alpha$ profile across the outburst event (Fig. \ref{fig:inset}), being the most remarkable examples those observed during the hard state (GTC-1 and GTC-2). In GTC-1 the profile is double-peaked with the red peak clearly stronger by a factor 2. The profile also shows a broad, blue wing extending down to $\sim -2000\, {\rm km \, s^{-1}}$ , while the red peak reaches the continuum level much more abruptly ($\sim 1000\, {\rm km \, s^{-1}}$). The observed peak-to-peak separation is $DP=1160 \pm  30\,\rm{km\, s^{-1}}$, and the profile is blue-shifted, with the centroid at $\gamma = -109 \pm  15\,\rm{km\, s^{-1}}$. During GTC-2, the profile shows an opposite behavior: the double peak is symmetric, while the peak-to-peak separation has significantly shrunk to $DP=753 \pm  10\,\rm{km\, s^{-1}}$, and the center of the 2-Gaussian structure is now red-shifted at $\gamma=173 \pm  9\,\rm{km\, s^{-1}}$. The striking difference between the two profiles is even clearer when we plot both epochs together (Fig. \ref{fig:overplot}). The two red peaks match each other, but the inter-peak depression on GTC-1 coincides with the blue peak in GTC-2. A P-Cygni profile dipping $\sim 2\%$ below the continuum level is also shown in GTC-2, with velocities of $\sim 1000\, {\rm km \, s^{-1}}$ and $\sim 1250\, {\rm km \, s^{-1}}$ for the core and its blue-edge velocity, respectively. The red wing extends now to $\sim 2200\, {\rm km \, s^{-1}}$. A weaker blue-shifted absorption is also observed in the nearby He \textsc{i} $6678.15\rm{\AA}$ line, albeit with a slightly lower blue-edge velocity of $\sim 1000\, {\rm km \, s^{-1}}$. 

Epochs GTC-3 to GTC-10, including VLT-1, show double-peaked profiles with peak-to-peak separation ($DP=500-700 \, {\rm km \, s^{-1}}$) and centroid ($\gamma=40-140 \, {\rm km \, s^{-1}}$) closer to that exhibited in GTC-2. They also show a broad absorption, where the emission line is embedded, and whose depth varies between epochs (with the deepest example being GTC-3, reaching $8\%$ below the continuum level), but always constrained within $\pm 2000 \, {\rm km \, s^{-1}}$. Similar broad absorptions have been previously observed in a handful of LMXBs (e.g., MAXI~J1807$+$132; \citealt{JimenezIbarra2019b}), but their origin is not well understood yet. \citet{Dubus2001} proposed they are produced in optically thick regions of the disc, mimicking the behavior of a stellar atmosphere. However, this scenario favors the observation of such features in low-inclination systems, which is at odds with their detection in relatively high inclination LMXBs such as XTE J1118+480 \citep{Khargharia2013}, and now J1803. The relative intensity of the red and blue emission peaks is also variable, partly due to the changing depth of the underlying broad absorption, but their height ratio ($H_{\rm red}/H_{\rm blue}=0.9-1.4$) is never comparable to that seen in GTC-1 ($H_{\rm red}/H_{\rm blue}=2.0$). The base of the emission component is restricted to the $\pm 1000\, {\rm km\, s^{-1}}$ range across all epochs, again remarking the striking difference with GTC-1 and GTC-2.

The latest epoch (GTC-11), taken three months after the outburst peak, shows a symmetric profile with larger double peak separation than in previous observations ($DP=980\pm 20\, {\rm km\, s^{-1}}$) but not as large as that observed in GTC-1. While this is still a soft state spectrum, the profile is broader than those observed at higher luminosity and similar to typical LMXB quiescent profiles. No obvious absorption component is present at this time.

\section{Discussion} \label{sec:discussion}

J1803 has shown X-ray dipping phenomenology \citep{Xu2021} similar to that found in other LMXBs (e.g. MAXI~J1305$-$704, \citealt{Shidatsu2013}). Such features are observed in high-inclination systems, a scenario further confirmed by the shape of the HID diagram presented in this paper. The periodicity of the dips serves as a good proxy for the true orbital period of the system. Therefore, we will hereafter consider $P_{\rm orb}\sim 7\, {\rm h}$, as well as a conservative lower limit on the orbital inclination of $i\gtrsim 65 \, {\rm deg}$.

\subsection{Optical accretion disk winds}

To date, up to six LMXBs have been found to exhibit P-Cygni profiles in their near-infrared and/or optical spectra: GX~13$+$1 \citep{Bandyopadhyay1997}, V404 Cyg \citep{Casares1991, Munoz-Darias2016, Munoz-Darias2017,MataSanchez2018}, V4641 Sgr \citep{Munoz-Darias2018}, Swift~J1858.6$-$0814 \citep{Munoz-Darias2020}, MAXI~J1820$+$070 \citep{Munoz-Darias2019} and GRS~1716$-$249 \citep{Cuneo2020b}. Our spectroscopy of J1803 showed P-Cygni profiles during GTC-2 in two emission lines (H$\rm \alpha$ and He \textsc{i} $6678\rm{\AA}$). The P-Cygni profiles have (blue-edge) terminal velocities of $1000-1250 \, {\rm km\, s^{-1}}$, which falls within the typical range observed in other LMXBs (\mbox{$\sim 1000- 3000\, {\rm km\, s^{-1}}$}). The detection of these optical wind features in J1803, a BH with a proposed orbital period as low as $P_{\rm orb}\sim 7\, {\rm h}$ \citep{Xu2021}, shows that these can develop over a wide range of accretion disk sizes (see e.g. \citealt{Munoz-Darias2020}).

Another significant feature traditionally associated with outflows is the presence of broad emission wings. These have been previously observed in a handful of LMXBs, sometimes simultaneously with the P-Cygni profiles (e.g. \citealt{MataSanchez2018}). They appear as either symmetric, broad wings, typically detected after a bright flare and accompanied by an enhanced Balmer decrement \citep{Munoz-Darias2016,MataSanchez2018}; or as a non-symmetric blue-shifted emission wing (e.g., MAXI~J1820$+$070; \citealt{SanchezSierras2020}). Regarding J1803, the most promising example is shown by the GTC-1 spectrum, with a blue wing extending to $\sim - 2000\, {\rm km\, s^{-1}}$, similar in absolute value to the red-wing of the GTC-2 P-Cygni spectrum. The H$\rm \alpha$ profile in GTC-1 is particularly unusual, as proven by its large full-width-at-half-maximum (FWHM) and $DP$, as well as its asymmetry. A possible interpretation involves the presence of a strong, blue-shifted absorption at low velocity ($\sim -500\, {\rm km\, s^{-1}}$) which cancels the blue-peak, thus producing the observed profile. Low-velocity P-Cygni profiles are intrinsically harder to identify, as they produce a dent on the broad emission from the accretion disk rather than a clear dip below the continuum. The combination of all these features in the GTC-1 J1803 spectrum leads us to conclude that outflows are present since the onset of the outburst, consistent with preliminary reports by \citet{Buckley2021}.

Finally, other possible outflow signatures include blue-shifted absorption components during optical dips (e.g., Swift~J1357.2$-$0933; \citealt{JimenezIbarra2019a}), broad wing emission components only detected at the end of the outburst (e.g. \citealt{Rahoui2014,Panizo2021}) and complex profiles such as flat-top or triangular profiles (MAXI~J1820$+$070; \citealt{SanchezSierras2020}). In this regard, VLT-1 reveals flat-top and triangular profile shapes in near-infrared lines such as Pa$\rm \beta$ or Pa$\rm \gamma$ (saw-tooth profile), but also flat-top lines like He \textsc{ii} $4686\, {\rm \AA}$ or Pa9. This leads us to propose that outflows are also present during the soft state, and are preferentially traced by the near-infrared spectrum. A similar conclusion was derived from the study of MAXI~J1820$+$070 \citep{SanchezSierras2020}. They suggested that this is likely the result of ionization effects rather than intrinsic changes in the outflow rate or geometry.

\subsection{On the nature of the compact object}

J1803 has been considered as a candidate to harbor a BH attending to the displayed X-ray properties during the outburst. However, only dynamical studies of the source during quiescence enable the determination of the compact object mass, thus confirming its nature. On this regard, \citet{Casares2015} introduced a new method to determine the companion star radial velocity semi-amplitude ($K_2$) from a scaling with the FWHM of the H$\rm \alpha$ emission line. Nevertheless, this correlation only holds for spectra obtained during true quiescence, a state occurring from months to years after the outburst (see e.g., \citealt{Casares2015}). While the study of this parameter during the earlier stages of the event is hampered by the complexity of the profiles (i.e., due to the presence of outflows), soft-state spectra are expected to evolve smoothly towards the quiescence level, as the disk shrinks to its original size. 

The H$\rm \alpha$ line during the J1803 outburst does indeed show significant variability in FWHM between $\sim 900-1600\, {\rm km\, s^{-1}}$, an expected behavior as the accretion disk expands. On this regard, our latest observed spectrum (GTC-11, 3 months after the outburst onset) shows the largest $\rm{FWHM}=1570\pm 100\, {\rm km\, s^{-1}}$, favoring this interpretation for J1803. In an attempt to assess the true quiescence FWHM from the aforementioned value, we inspected the literature to assess typical quiescence to outburst correction ratios ($\rm{FWHM}_{\rm ratio}=\rm{FWHM}_{\rm quies}$/$\rm{FWHM}_{\rm out}$). We note that larger corrections are required if hard-state or early soft-state epochs are considered, due to a combination of a smaller accretion disk and the influence of outflows. For this reason, we focused on optical spectra at low X-ray luminosity and as late in the event as possible (such as our J1803 GTC-11 epoch). We found the best comparison with the recent outburst of MAXI~J1820$+$070, a high-inclination transient BH for which we take $\rm{FWHM}_{\rm quies}=1793 \pm 101\, {\rm km\, s^{-1}}$ \citep{Torres2020} and a range of $\rm{FWHM}_{\rm out}=1180-1360\, {\rm km\, s^{-1}}$, from measurements in \citet{Munoz-Darias2019} outburst spectra of $1220\pm 40\, {\rm km\, s^{-1}}$ (epoch 32) and $1340\pm 20\, {\rm km\, s^{-1}}$ (epoch 33). This corresponds to a conservative range of $\rm{FWHM}_{\rm ratio}\sim 1.2-1.6$, consistent with that observed in other transient BHs, e.g. $\rm{FWHM}_{\rm ratio}\sim 1.2-1.3$ for Swift~J1357.2$-$0933 \citep{Torres2015}, and $\rm{FWHM}_{\rm ratio}\sim 1.4$ for V404 Cyg \citep{MataSanchez2018}. 

We use the above estimate to construct a uniform distribution of $\rm{FWHM}_{\rm ratio} $ to be combined with a normal distribution for the measured $\rm{FWHM}$, yielding $\rm{FWHM}_{\rm quies}\sim 2000-2500 \, {\rm km\, s^{-1}}\, (1800-2700 \, {\rm km\, s^{-1}})$ at the $68\%$ ($95\%$) confidence level. Following \citet{Casares2015}, we derive $K_2\sim 460-570 \, {\rm km\, s^{-1}}\, (410-620 \, {\rm km\, s^{-1}})$. This, together with $P_{\rm orb}\sim 7\, {\rm h}$, yields a lower limit to the mass of the compact object of $\gtrsim 3\, M_{\rm \odot}$, providing further support to its BH nature. In addition, by considering the constraints on the orbital inclination ($i>65\, {\rm deg}$) and assuming a typical range for the binary mass ratio ($0.01\leq q \leq0.2$) we obtain $M_{\rm BH}\sim 4-8\, M_{\rm \odot} (3-10\, M_{\rm \odot})$. This mass range is consistent with that of the known population of BHs in LMXBs \citep{Casares2014}.

\section{Conclusions}

We presented a multi-wavelength follow-up of the discovery outburst of the BH candidate MAXI~J1803$-$298. We detected P-Cygni profiles in H$\rm \alpha$ and He \textsc{i} lines, with a terminal velocity of $\sim 1250\, {\rm km\, s^{-1}}$. We also detected a particularly broad emission line blue wing, together with an apparent peak-to-peak separation much larger than observed during the rest of the outburst. All these features are seen during the initial hard state, indicating the presence of a wind-type outflow during this stage of the outburst. The soft-state spectra are instead characterized by narrower emission lines embedded into broad absorption components. On the other hand, the near-infrared emission lines show more complex profiles, that can also be tentatively associated with outflows. Finally, we provide further support for the identification of the compact object in J1803 as a BH, and propose a mass range of $M_{\rm BH}\sim 3-10\, M_{\rm \odot}$.

\begin{acknowledgments}
DMS and MAP acknowledge support from the Consejer\'ia de Econom\'ia, Conocimiento y Empleo del Gobierno de Canarias and the European Regional Development Fund (ERDF) under grant with reference ProID2020010104 and ProID2021010132. TMD and MAPT acknowledge support via Ram\'on y Cajal Fellowships RYC-2015-18148 and RYC-2015-17854, respectively. This work has been supported in part by the Spanish Ministry of Science under grants AYA2017-83216-P, PID2020-120323GB-I00 and EUR2021-122010. We thank Tom Marsh for the use of \textsc{molly} software. We are thankful to the GTC staff for their prompt and efficient response at triggering the time-of-opportunity program at the source of the spectroscopy presented in this Letter. Based on observations collected at the European Southern Observatory under ESO programme 105.20LK.002.
\end{acknowledgments}

%

\vspace{5mm}
\facilities{GTC(OSIRIS), VLT(XSHOOTER), NICER, MAXI(GSC)}


\software{astropy \citep{2013A&A...558A..33A}, 
          \textsc{iraf} \citep{iraf1986},
          \textsc{molly}( \url{http://deneb.astro.warwick.ac.uk/phsaap/software/molly/html/INDEX.html}),
          \textsc{pyraf}( \url{https://iraf-community.github.io/pyraf.html}, \url{http://ascl.net/1207.011}),
          \textsc{xspec} (v.12.12.0, \citealt{Arnaud1996}),
          }


\bibliography{bibliography}{}
\bibliographystyle{aasjournal}



\end{document}